\documentclass[12pt]{article}
\usepackage{times}
\usepackage{amsmath}
\usepackage{amssymb}
\usepackage{amsthm}
\usepackage{booktabs}
\usepackage{graphicx}
\usepackage{hyperref}
\usepackage{subfigure}
\usepackage[super,comma]{natbib}
\usepackage{setspace} 
\usepackage{refstyle}
\usepackage{nameref} 
\newcommand{\newnameref}[1]{\emph{\nameref{#1}}} 
\makeatletter
\newcommand{\labelname}[1]{
  \def\@currentlabelname{#1}}%
\makeatother
\usepackage[hmargin=1.1in,vmargin=1.1in]{geometry}
\usepackage{soul}

\usepackage{etoolbox}
 
\makeatletter
\patchcmd{\maketitle}{\@fnsymbol}{\@alph}{}{}  
\makeatother

\begin{document}

\title{Bracketing in the Comparative Interrupted Time-Series Design to Address Concerns about History Interacting with Group: Evaluating Missouri Handgun Purchaser Law}
\author{Raiden B. Hasegawa\thanks{Department of Statistics, The Wharton School, University of Pennsylvania, Philadelphia, PA}\\Daniel W. Webster\thanks{Department of Health Policy and Management, Johns Hopkins Bloomberg School of Public Health, Johns Hopkins University, Baltimore, MD.}\\Dylan S. Small \thanks{Correspondence: Dylan S. Small, Department of Statistics, The Wharton School, University of Pennsylvania, Philadelphia, PA. (E-mail: {\it{dsmall@wharton.upenn.edu}})}}
\date{\today}
\maketitle

\thispagestyle{empty}





{\it{Data and code availability}}: The data is provided in Table 1.  The code that produced the

results can be found at the authors website \url{http://www.raidenhasegawa.com}. 

{\it{Keywords}}: Causal inference; bracketing; difference-in-difference; 
comparative interrupted 

time series; permit-to-purchase; history-by-group interaction; unmeasured confounding; 

multiple control groups; gun violence; firearm policy
 
\newpage 

\begin{abstract}
In the comparative interrupted time series design (also called the method of difference-in-differences), the change in outcome in a group exposed to treatment in the periods before and after the exposure is compared to the change in outcome in a control group not exposed to treatment in either period.  The standard difference-in-difference estimator for a comparative interrupted time series design will be biased for estimating the causal effect of the treatment if there is an interaction between history in the after period and the groups; for example, there is a historical event besides the start of the treatment in the after period that benefits the treated group more than the control group.  We present a bracketing method for bounding the effect of an interaction between history and the groups that arises from a time-invariant unmeasured confounder having a different effect in the after period than the before period.  The method is applied to a study of the effect of the repeal of Missouri's permit-to-purchase  handgun law on its firearm homicide rate. We estimate that the effect of the permit-to-purchase repeal on Missouri's firearm homicide rate is bracketed between 0.9 and 1.3 homicides per 100,000 people, corresponding to a percentage increase of 17\% to 27\% (95\% confidence interval: [0.6,1.7] or [11\%,35\%]). A placebo study provides additional support for the hypothesis that the repeal has a causal effect of increasing the rate of state-wide firearm homicides.
\end{abstract}

\section*{Comparative Interrupted Time Series Design and Potential Biases}

\begin{figure}[h!]
    
    \caption{Stylized plot of data from a comparative interrupted time series design.  The dotted line shows the assumption that the difference-in-difference (DiD) estimate makes about the treatment group'€™s counterfactual mean in the absence of treatment.}\label{parallel.trends}
    \centering
    \vspace{0.5cm}
    \includegraphics[scale=0.5]{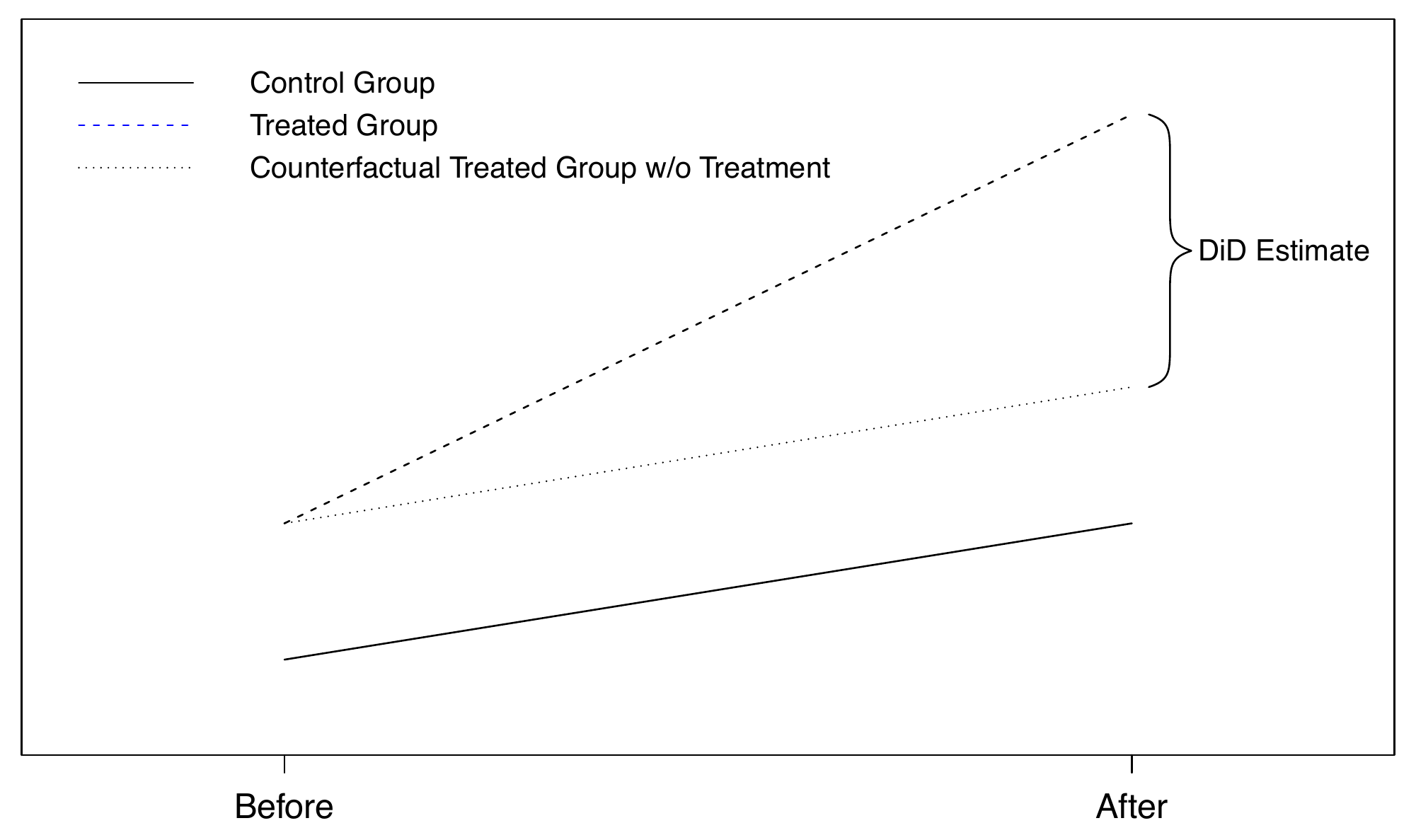}
\end{figure}

The interrupted time series design is an observational study design for estimating the causal effect of a treatment on a group when data is available before the group was treated.
In the simplest interrupted time series design, the before and after treatment outcomes are compared.  This before-after design does not account for confounding factors that co-occur with treatment such as historical events or maturation\citep{cook2002experimental}.  To strengthen the before-after design, it is common to add time series data from a control group that never received the treatment over the same period {\textendash} the comparative interrupted time series  design\citep{cook2002experimental,meyer1995natural,bernal2017interrupted,wing2018designing}, also called the nonequivalent control group design or method of difference-in-differences.  The latter name derives from the concept that the simplest comparative interrupted time series analysis is to take the difference between the difference of the after and before outcomes for the treated group and the difference of the after and before outcomes for the control group.  This difference-in-differences estimate is an unbiased estimator of the causal effect of treatment if the treatment and control groups would have exhibited parallel trends in the counterfactual absence of treatment\citep{meyer1995natural}; see Figure \ref{parallel.trends}.  
The parallel trends assumption can be partially assessed if there is more than one time point in the before period by assessing whether the groups exhibit parallel trends in the before period \citep{meyer1995natural}.  However, even if the trends are parallel in the before period, there could be historical events in the after period that affect the two groups differently, i.e., history interacts with group (other reasons that parallel trends could be violated include differences in maturation, instrumentation or statistical regression between the groups)\citep{cook1979,reynolds1987multiplist}.  For example, the outcome measures poor health, country $A$ (treated group) enacts a policy reform, country $B$ (control group) does not enact the reform, and a worldwide economic recession occurs after the reform that has a greater impact on people starting out in poorer health.  If country $B$ started out with poorer health, then parallel trends would be violated because country $B$'s poor health would have increased more than country $A$ in the after period in the counterfactual absence of the reform because of the worldwide economic recession.  This violation of parallel trends would not happen if $A$ and $B$ started with the same level of poor health in the before period.  However, it is often difficult to find a control group that has outcomes close to the treated group in the before period.

When there is no control group completely comparable to the treated group, \citet{campbell1969} proposed bracketing to distinguish treatment effects from plausible biases\citep{rosenbaum1987}. For comparing treatment and control at one time point, suppose there is concern about an unmeasured confounder $U$.  Bracketing uses two control groups such that, in the first group $U$ tends to be higher than in the treated group and in the second group, $U$ tends to be lower.  The effect of $U$ on the treated group is bracketed by its effect on the two control groups.  When there is bracketing, if the treated group has a notably higher outcome than both control groups, then this association between treatment and outcome cannot plausibly be explained away as being bias from $U$.

In this paper, we show how bracketing can be applied to the comparative interrupted time series to distinguish treatment effects from plausible biases due to history interacting with group.
The basic idea is to consider one control group that has a lower expected outcome than the treated group in the before period and another control group that has a higher expected outcome than the treated group in the before period; we show under certain assumptions that the expectations of the two difference-in-difference estimators using the lower control group and higher control group respectively bracket the causal effect of the treatment.  Bracketing for the comparative interrupted time series has been mentioned informally \citep{meyer1995natural} but the idea of choosing the bracketing control groups based on expected before period outcomes was not mentioned.  We present assumptions and results for our bracketing method (\newnameref{sec:bracket}) and then apply the method to study the effect of the repeal of Missouri's permit-to-purchase handgun law on its firearm homicide rate (\newnameref{sec:application}).

\section*{Methods: Bracketing}\label{sec:bracket}

\subsection*{Notation and Model}
\label{subsec:notation.model}

Let $Y$ denote outcome and $D$ dose of exposure, $D=1$ for treatment and $D=0$ for control.
Let $Y_{ip}^{(d)}$ denote the counterfactual outcome that would have been observed for unit $i$ in period $p$, $p=0$ for before period and $p=1$ for after period, had the unit received exposure dose $d$, i.e., $Y_{ip}^{(1)}$ is the counterfactual outcome under treatment and $Y_{ip}^{(0)}$ is the counterfactual outcome under control.  Let ${\bf{U}}_i$ be a vector of time invariant unmeasured confounders for unit $i$.  Let $G$ denote group where the groups are $t=$ treated group, $lc=$ lower control group (control group with expected outcomes lower than treated group in before period) and $uc=$ upper control group (control group with expected outcomes higher than treated group in before period).  Finally, let $S$ be an indicator of whether or not a unit belonging to a particular group is in the study population in a given period. Specifically, $S_{ip}=1$ or $0$ when unit $i$ is in the population or not in period $p$: $S_{i0}=S_{i1}=1$ for a unit in the population both before and after treatment, $S_{i0}=1,S_{i1}=0$ for a unit in the population only before treatment (unit might have moved away or died in after period) and $S_{i0}=0,S_{i1}=1$ for a unit in the population only after treatment (unit might have moved into study area or been born in after period).

We consider the following model which generalizes the standard difference-in-difference model and changes-in-changes model.\citep{athey2006identification}  Let ${\bf{U}}_i$ be time-invariant unmeasured confounders and $\epsilon_{ip}$ be an error term that captures additional sources of variation for unit $i$ in period $p$.  Then our model can be expressed as 
\begin{equation}
    Y_{ip}^{(d)}=h({\bf{U}}_i,p)+\beta d + \epsilon_{ip} \label{modeling.statement}
\end{equation}
where the function $h({\bf U}_i,p)$ is the unobserved expected outcome under control of subject $i$ in period $p$. We drop the subscript $i$ to refer to a randomly drawn unit from the population of all units in either period, where $Y_p^{(d)}, d=0,1,$ and $\epsilon_p$ are undefined if $S_p=0$. We make the following assumptions:
\begin{gather}
\mbox{\textit{Increasingness of $h$ in ${\bf{U}}$}: $h({\bf{U}},p)$ bounded and increasing in ${\bf{U}}$ for $p=0,1$}. \label{increasingness.in.u} \\
\mbox{($(h({\bf{U}},p)\geq h({\bf{U}}^{'},p)$ whenever all coordinates of ${\bf{U}}$ $\geq$ all coordinates of ${\bf{U}}^{'}$)} \nonumber \\
\nonumber\\
\mbox{{\it Time Invariance of ${\bf{U}}$ Within Groups}: ${\bf{U}}$ conditionally independent of}\label{time.invariance}\\ 
\mbox{$\{ S_0,S_1\}$ given group $G$.}  \nonumber \\
\nonumber\\
\mbox{\textit{Independence of $\epsilon$ with Time and Group}: Distributions of $\epsilon_{p}|S_p=1,G=g$ for} \label{epsilon.invariance}\\
\mbox{$p=0,1$, $g=lc,uc,tc$ all have mean zero and are the same.} \nonumber
\end{gather}
Assumptions (\ref{increasingness.in.u}) and (\ref{time.invariance}) match assumptions in the changes-in-changes model.  Assumption (\ref{increasingness.in.u}) requires that higher levels of unmeasured confounders correspond to higher levels of outcomes.  Such increasingness is natural when the unmeasured confounder is an individual characteristic such as health or ability \citep{athey2006identification} and $Y$ is a measure of some positive outcome, for example, income. Negative confounders -- where higher levels of the confounder correspond to lower levels of the outcome -- are not precluded by Assumption (\ref{increasingness.in.u}) as the corresponding coordinates of $\bf{U}$ may simply be replaced by their negation. Assumption (\ref{time.invariance}) says that the distribution of confounders in the population of units for a given group remains the same over time.   Assumption (\ref{epsilon.invariance}) says that time-varying factors have the same distribution in each group and over time. It would be sufficient for subsequent developments to just assume the distributions of $\epsilon_{p}|S_p=1,G=g$ for $p=0,1$, $g=lc,uc,tc$ all have mean zero rather than the stronger assumption of identical distributions.  We can further relax this assumption by assuming zero mean only for components of $\epsilon_p$ that are true confounders, that is, factors whose distributions depend on the interaction of time and group. Assumption (\ref{epsilon.invariance}) is weaker than the changes-in-changes model assumption that $\epsilon_{ip}$ is always zero which rules out classical measurement error in the outcome when $h$ is non-linear.\citep{athey2006identification}  Our model contains the standard difference-in-difference model, which can be represented in our model by $h({\bf{U}},p)=k({\bf{U}})+\tau p$ for some bounded and increasing function $k$, where $k({\bf U})$ can be viewed as a group fixed effect.  

We make two further assumptions about the distribution of ${\bf{U}}$ in groups and how its effect over time changes among the groups.  First, we assume the distribution of ${\bf{U}}$ within groups can be stochastically ordered so that ${\bf{U}}$ is lowest in the lower control group, intermediate in the treated group and highest in the upper control group:
\begin{equation}
{\bf{U}}|G=lc \preceq {\bf{U}}|G=t \preceq {\bf{U}}|G=uc 
\label{stochastic.dominance}
\end{equation}
where two random vectors ${\bf{A}},{\bf{B}}$ are stochastically ordered, ${\bf{A}}\preceq {\bf{B}}$, if $E[f({\bf{A}})]\leq E[f({\bf{B}})]$ for all bounded increasing functions $f$ \citep{shakedstochastic}. For example, if $\bf{U}$ is normally distributed with common variance and group means $\mu_{lc},\mu_{t},$ and $\mu_{uc}$, then  $\mu_{lc}\le \mu_{t}\le \mu_{uc}$ would imply (\ref{stochastic.dominance}).
Second, we assume that higher values of ${\bf{U}}$ either have a bigger effect over time over the whole range of {$\bf{U}$ or a smaller effect over the whole range:
\begin{align}
\mbox{Either }&\mbox{ (i) $h({\bf{U}},1)-h({\bf{U}},0)\geq h({\bf{U}}^{'},1)-h({\bf{U}}^{'},0)$ for all ${\bf{U}}\geq {\bf{U}}^{'}$, ${\bf{U}},{\bf{U}}^{'}\in {\mathcal{U}}$ }\mbox{ or} \nonumber \\
&\mbox{ (ii) $h({\bf{U}},1)-h({\bf{U}},0)\leq h({\bf{U}}^{'},1)-h({\bf{U}}^{'},0)$ for all ${\bf{U}}\geq {\bf{U}}^{'}$, ${\bf{U}},{\bf{U}}^{'}\in {\mathcal{U}}$}
 \label{increasing.differences}
\end{align}

An example of this pattern of ${\bf{U}}$ confounding could occur in a study of the effect of a regional policy on average income where the policy change occurred contemporaneously with an easing of trade restrictions.  A potential unmeasured confounder for such a study would be ${\bf{U}}$ = share of skilled workers in a region, as a higher share of skilled workers is associated with higher average income.  There is considerable evidence that trade liberalization leads to an increase in the skill premium -- the relative wage of skilled to unskilled workers -- at both the regional and country level \citep{dix2017trade,burstein2017international}.  Thus, we might expect (i) in (\ref{increasing.differences}) to hold if there was an easing of trade restrictions in the after period.   

We assume units are randomly sampled from each group in each time period. The data could be obtained from repeated cross sections or a longitudinal study. Inferences under different sampling assumptions are discussed in \nameref{eAppendix1}.

\subsection*{Bracketing Result}
\label{subsec:bracketing}
The standard moment difference-in-difference estimator using control condition $c$ can be written as $\hat\beta_{dd.c} = (\overline{Y}_{1|G=t} - \overline{Y}_{0|G=t}) - (\overline{Y}_{1|G=c} - \overline{Y}_{0|G=c})$ where $\overline{Y}_{p|G=g}$ indicates the sample average of units observed in group $g$ and time period $p$, $Y_p|G=g,S_p=1$. This estimate is equivalent to the coefficient on the treatment indicator in a fixed effects regression with full time and group indicator variables. When using data already aggregated at some level, for example by state-year, a fixed effects regression using weights proportional to population will return this estimate. In the following, we show that the expectation of the two standard difference-in-difference estimators computed with the upper and lower controls can be used to bound the treatment effect.

The expected value of the standard difference-in-difference estimator comparing the treated group to the lower control group, $\hat{\beta}_{dd.lc}$, is
\begin{eqnarray*}
E[\hat{\beta}_{dd.lc}] & = & \{ E[Y_1|G=t,S_1=1]-E[Y_0|G=t,S_0=1]\} \\
& & - \{ E[Y_1|G=lc,S_1=1]-E[Y_0|G=lc,S_0=1]\} \\
& = & \{\beta+E[h({\bf{U}},1)|G=t,S_1=1]-E[h({\bf{U}},0)|G=t,S_0=1]\} \\
& & - \{E[h({\bf{U}},1)|G=lc,S_1=1]-E[h({\bf{U}},0)|G=lc,S_0=1]\},
\end{eqnarray*}
where $Y_1,Y_0$ denote observed outcomes in after period ($p=1$) and before period ($p=0$) respectively.
Under the time invariance of ${\bf{U}}$ within groups assumption (\ref{time.invariance}), we have
\begin{equation}
E[\hat{\beta}_{dd.lc}]=\beta+\{E[h({\bf{U}},1)-h({\bf{U}},0)|G=t]\} - \{E[h({\bf{U}},1)-h({\bf{U}},0)|G=lc]\}; \label{diff.in.diff.bias.lc}
\end{equation}
similarly, the expected value of the difference-in-difference estimator comparing the treated group to the upper control group, $\hat{\beta}_{dd.uc}$, is
\begin{equation}
E[\hat{\beta}_{dd.uc}]=\beta+\{E[h({\bf{U}},1)-h({\bf{U}},0)|G=t]\} - \{E[h({\bf{U}},1)-h({\bf{U}},0)|G=uc]\}. \label{diff.in.diff.bias.uc}
\end{equation}
The difference-in-difference estimators $\hat{\beta}_{dd.lc}$ and $\hat{\beta}_{dd.uc}$ are unbiased if $h({\bf{U}},1)-h({\bf{U}},0)$ is constant for all ${\bf{U}}$, i.e., the time effect between periods is the same for all levels of $\bf{U}$, or equivalently, the effect of the unmeasured confounders is the same in both time periods. If the effect of the unmeasured confounders changes between periods, then because of assumptions (\ref{stochastic.dominance}) and (\ref{increasing.differences}), we conclude from (\ref{diff.in.diff.bias.lc}) and (\ref{diff.in.diff.bias.uc}) that
\begin{equation}
\min \{ E[\hat{\beta}_{dd.lc}],E[\hat{\beta}_{dd.uc}]\} \leq \beta \leq \max \{ E[\hat{\beta}_{dd.lc}],E[\hat{\beta}_{dd.uc}]\}, \label{bracketing.result}
\end{equation}
i.e., the expected values of the difference-in-difference estimators using the upper control group and lower control group bracket the causal effect (proof in \nameref{eAppendix2}). The tightness of the bracketing bounds in (\ref{bracketing.result}) and, to some extent, the width of the corresponding confidence interval developed in following section depend on the magnitude of the group-by-time interaction. For example, if urban poverty concentration varied notably between groups and its effect on firearm homicides were modulated by the Great Recession, one would expect looser bracketing bounds. 

\subsection*{Inference}
\label{subsec:inference}

We would like to make inferences for the causal effect $\beta$ under the assumption (\ref{increasing.differences}) that $h({\bf{U}},1)-h({\bf{U}},0)$ is either an increasing or decreasing function of ${\bf{U}}$ (we do not want to specify which a priori).  Let $\theta_{lc.t}=E[\hat{\beta}_{dd.lc}]$ and  $\theta_{uc.t}=E[\hat{\beta}_{dd.uc}]$, i.e., the expected values of the difference-in-difference estimators using the lower control group and upper control group, respectively.  From the bracketing results (\ref{bracketing.result}), we have
\[
\min (\theta_{lc.t},\theta_{uc.t}) \leq \beta \leq \max (\theta_{lc.t},\theta_{uc.t}).
\]
and the following interval, where CI means confidence interval,
\begin{gather}
\mbox{[min(lower endpoint of $1-\alpha$ two sided CI for $\theta_{lc.t}$, lower endpoint of $1-\alpha$ two sided CI for $\theta_{uc.t}$),} \nonumber \\
\mbox{max(upper endpoint of $1-\alpha$ two sided CI for $\theta_{lc.t}$, upper endpoint of $1-\alpha$ two sided CI for $\theta_{uc.t}$)]}, \label{minmax.ci}
\end{gather}
has probability $\geq 1-\alpha$ of containing both $\min (\theta_{lc.t},\theta_{uc.t})$ and $\max (\theta_{lc.t},\theta_{uc.t})$, and thus $\beta$, where it assumed that the two-sided CIs are constructed by taking the intersection of two one-sided $1-(\alpha /2)$ confidence intervals (proof in \nameref{eAppendix3}).  

\subsection*{Constructing the Lower and Upper Control Groups}
\label{subsec:constructing.groups}

The results in \newnameref{subsec:bracketing} and \newnameref{subsec:inference} assume the lower and upper control groups have been constructed before looking at the data.  If the lower control group was constructed by looking at the before period data by choosing units with lower outcomes than the treated in the before period, then the sample average of $Y_0|G=lc,S_0=1$ may tend to be lower than $E(Y_0|G=lc,S_0=1)$. Consequently, the difference-in-difference estimate using the lower control group may be downward biased even if the parallel trends assumption holds because of regression to the mean\citep{cook2002experimental}; similarly, the difference-in-difference estimated using the upper control group may be upward biased.  This may invalidate the bracketing result (\ref{bracketing.result}).  To avoid bias arising from regression to the mean, we propose first selecting a ``pre-study'' time period prior to the before period. Then, the lower control group can be constructed from units with lower outcomes than the treated in this pre-study period and the upper control group from units with higher outcomes.  It should then be tested whether the constructed lower control group has smaller expected outcomes than the constructed upper control group in the before period; see \newnameref{sec:application} for example.

\subsection*{Role of Examining the Groups' Relative Trends in the Before Period}\label{subsec:rel.trends}

In the standard difference-in-difference analysis that assumes parallel trends, when the before period contains multiple time points, it is good practice to test for parallel trends in the before period\citep{meyer1995natural,volpp2007mortality}.
In our bracketing approach, we do not need the parallel trend assumption to hold, but examining the relative trends of the groups in the before period is still useful for assessing  model plausibility and assumptions.  Our model (\ref{modeling.statement})-(\ref{epsilon.invariance}) along with assumptions (\ref{stochastic.dominance})-(\ref{increasing.differences}) implies that if we had counterfactual data on the treatment group in the after period in the absence of treatment, then, without sampling variance, we would see either: (i) the differences between the upper control and counterfactual treated groups and the difference between the counterfactual treated and lower control groups in the after period would be at least as large as their respective differences in the before period or (ii) the difference between the upper control and counterfactual treated groups and the difference between the counterfactual treated and lower control groups in the after period would be no larger and possibly smaller than their respective differences in the before period.  The following two patterns would violate the model/assumptions: (iii) the difference between the upper control and counterfactual treated groups is larger after than before and the difference between the counterfactual treated and lower control groups is smaller after than before or (iv) the difference between the upper control and counterfactual treated groups is smaller after than before and the difference between the counterfactual treated and lower control groups is larger after than before.  Although we do not have the counterfactual treatment group's data in the absence of treatment in the after period, we have the treatment group's data in the absence of treatment in the before period. We can split the before period into two (or more) periods and test whether the pattern in the before period is consistent with the model. Visual inspection of the relative trends of the counterfactual treated group and the upper and lower control groups during the before period can provide additional evidence for or against the model assumptions.

\subsection*{Time-Varying Confounders}
\label{subsec.time.varying.confounders}

Our bracketing method addresses an interaction between history and groups that arises because the time-invariant unmeasured confounders that differ between the groups in the before period (${\bf{U}}$) become more (or less) important in the after period (assumption (\ref{increasing.differences})).
 When there are time-varying confounders, the bracketing method still works under certain assumptions.  Time-varying confounders can be represented in model (\ref{modeling.statement}) by letting ${\bf{U}}$ contain all variables that differ in distribution between the groups in the before period, $\epsilon_{i0}$ be the effect of factors that do not differ in distribution between the groups in the before period and $\epsilon_{i1}$ be the effect of the same factors in $\epsilon_{i0}$ in the after period as well as factors not contained in ${\bf{U}}$ that differ in distribution between the groups in the after period (details on time-varying model in \nameref{eAppendix4}).  If this last set of factors is present, then (\ref{epsilon.invariance}) may not hold.  However, the bracketing result (\ref{bracketing.result}) still holds as long as (i) in (\ref{increasing.differences}) holds,
 \begin{equation}
    E[\epsilon_{i1}|G=uc]\geq E[\epsilon_{i1}|G=t]\geq E[\epsilon_{i1}|G=lc], \label{time.varying.confounders.assumption.i}
\end{equation}
or when (ii) in (\ref{increasing.differences}) holds,
\begin{equation}
    E[\epsilon_{i1}|G=uc]\leq E[\epsilon_{i1}|G=t]\leq E[\epsilon_{i1}|G=lc]; \label{time.varying.confounders.assumption.ii}
\end{equation}
\nameref{eAppendix4} contains a proof and sufficient conditions for (\ref{time.varying.confounders.assumption.i}) or (\ref{time.varying.confounders.assumption.ii}) to hold.  One of these sufficient conditions (condition (c) in \nameref{eAppendix4}) is analogous to (i) in (\ref{increasing.differences}) in that effects on the outcome, be they time effects or those due to contemporaneous shocks to confounders, are amplified at larger values of $\mathbf{U}$.

One type of time-varying confounder is a variable that largely stays the same between time periods but may change modestly.  For example, in our study of Missouri's repeal of their permit-to-purchase law in \newnameref{sec:application}, urban concentration of poverty might be a confounder and ${\bf{U}}$ contain urban concentration of poverty in the before period.  Urban concentration of poverty may stay mostly the same over time but change modestly, where the changes are reflected in $\epsilon_1$.  If the effect of urban concentration of poverty on firearm homicides increased in the after period, then the bracketing result would still hold (with respect to the confounding from urban concentration of poverty) as long as the impact of changes in urban concentration of poverty on firearm homicides were at least as great in the upper control group as Missouri and at least as great in Missouri as the lower control group.  

\section*{Application: Effect of the Repeal of Missouri's Handgun Purchaser Licensing Law on Firearm Homicides}
\label{sec:application}

 American federal gun law requires background checks and record keeping for gun sales by federally licensed firearm dealers but exempts these regulations for private sales.  However, some states have laws requiring all purchasers of handguns from licensed dealers {\it and} private sellers to acquire a permit-to-purchase  license that verifies the purchaser has passed a background check.  Missouri passed a permit-to-purchase law in 1921, requiring handgun purchasers to obtain a license from the local sheriff's office that facilitated the background check, but repealed the law on August 28, 2007.  \citet{webster2014effects} examined the effect of Missouri's repeal on firearm homicide rates (the rate of homicides committed using a firearm).  One of their analyses used a comparative interrupted time series design, comparing Missouri to the eight states bordering Missouri using a before-period of 1999-2007 and after-period of 2008-2010 (the only available post-repeal data at the time of their analysis), finding evidence that the repeal of Missouri's permit-to-purchase law increased firearm homicide rates (see their Table 1). None of the border states introduced new or made changes to existing permit-to-purchase laws during the study period. Using a fixed effect regression and adjusting for several background crime and economic covariates, they estimated that the Missouri permit-to-purchase repeal was associated with an increase in the firearm homicide rate by 1.1 per 100,000 persons (95\% confidence interval [CI]: 0.8,1.4) , a 22\% (95\% CI: 16 \%, 29\%) increase. Non-gun related homicides remained virtually unchanged. In what follows, we re-examine the effect of Missouri's repeal using bracketing and the now available after-period data from 2008-2016 to address possible biases arising from unobserved state-by-time interactions.

Figure \ref{border.trends} shows the age-adjusted firearm homicide rates in Missouri and the border states over the study period using data from the Centers for Disease Control and Prevention's (CDC) Wide-ranging Online Data for Epidemiologic Research (WONDER) system\citep{CDC2018}. The standard difference-in-difference estimate using all neighboring control states, shown in the top row of Table \ref{results.table}, is that Missouri's permit-to-purchase repeal increased firearm homicides by 1.2 per 100,000 persons (95\% CI: 1.0,1.4), corresponding to a 24\% increase (95\% CI: 18\%,31\%). 
\begin{figure}[h!]
    
    \caption{Age-adjusted firearm homicide rates in Missouri and states bordering Missouri (population-weighted averages), 1999-2016.}\label{border.trends}
    \centering
    \vspace{.5cm}
    \includegraphics[scale=0.6]{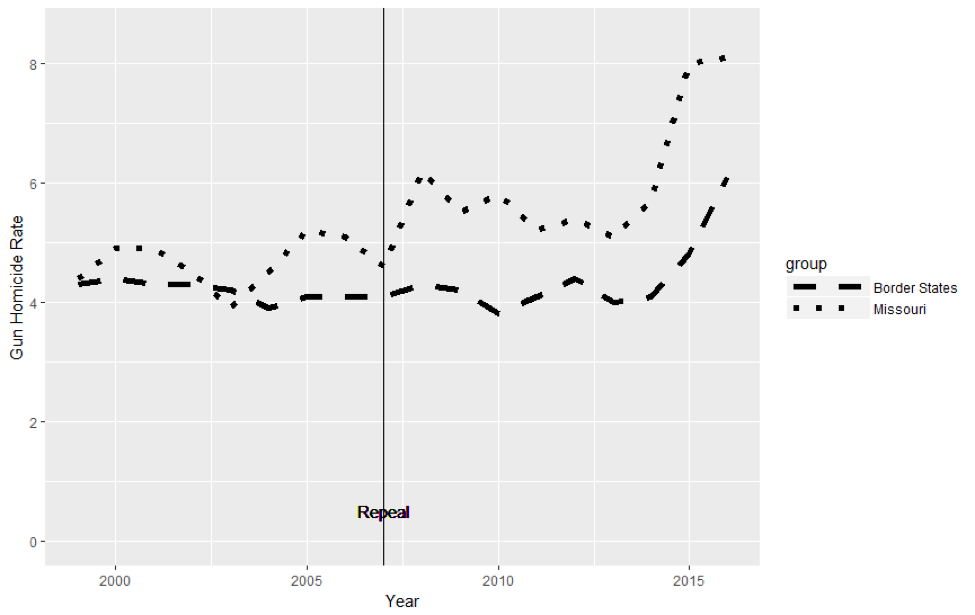}
\end{figure}
In the before-period, Missouri had generally higher firearm homicide rates than the control border states, suggesting a lack of comparability between the groups.  One concern is that the start of the after period coincided with the beginning of the Great Recession. The economic downturn was followed by a decline in homicide rates. Possible reasons for the effect of the downturn on homicide rates and violence generally include changing alcohol affordability, disposable income, unemployment, and income inequality\citep{matthews2006violence,wolf2014violence,shepherd2015economic}.  The effects of the economic downturn on firearm homicides might interact with the starting level of firearm homicides in a state.  To address this concern, we constructed upper and lower control groups that bracket Missouri's firearm homicide rate in the before period.  To avoid regression to the mean (\newnameref{subsec:constructing.groups}), we use data from 1994-1998, the five years prior to our before period, to choose the upper and lower control groups; see Table \ref{datatable} for data.  The lower control group is Iowa, Kansas, Kentucky, Nebraska, and Oklahoma and the upper control group is Arkansas, Illinois, and Tennessee.  The population-weighted firearm homicide rate in the before period of 1999-2007 is 5.2 in the upper control states, 4.7 in Missouri, and 2.7 in the lower control states (95\% CI for difference between upper control and Missouri: 0.2,0.8; 95\% CI for difference between Missouri and lower controls: 1.8,2.2).

\begin{table}[h!]
\caption{Age-adjusted firearm homicide rates per 100,000 persons from periods 1994-1998 (pre-study period used to construct lower and upper control groups), 1999-2007 (before repeal period where repeal refers to repeal of Missouri's permit-to-purchase handgun licensing law) and 2008-2016 (after repeal period).}
\begin{center}
\begin{tabular}{|l|c|c|c|}
\hline & 1994-1998 & 1999-2007 & 2008-2016 \\ \hline
Missouri & 6.1 & 4.7 & 6.1 \\ \hline
Arkansas & 7.3 & 5.1 & 5.5 \\
Illinois & 7.1 & 5.1 & 5.2 \\
Iowa & 1.2 & 0.9 & 1.2 \\
Kansas & 4.2 & 3.0 & 3.0 \\
Kentucky & 4.1 & 3.3 & 3.7 \\
Nebraska & 2.2 & 1.8 & 2.4 \\
Oklahoma & 4.8 & 3.8 & 4.8 \\
Tennessee & 6.9 & 5.5 & 5.4 \\ \hline
Population-weighted & 5.6 & 4.2 & 4.4 \\
\hspace{5mm} All Controls & & & \\
Population-weighted & 7.1 & 5.2 & 5.3 \\
\hspace{5mm} Upper Controls & & & \\
Population-weighted & 3.5 & 2.7 & 3.2 \\
\hspace{5mm} Lower Controls & & & \\ \hline
\end{tabular}
\end{center}
\label{datatable}
\end{table}
 
Figure \ref{bracket.trends} shows firearm homicides rates (age-adjusted and population-weighted) in the bracketed control groups compared to Missouri.  The bottom two rows of Table \ref{results.table} show the difference-in-difference estimates using the lower and upper control groups and 95\% CIs.  Both the lower and upper control groups provide evidence that Missouri's repeal of its permit-to-purchase handgun law increased firearm homicides, bracketing the effect of the repeal between 0.9 and 1.3 homicides per 100,000 people, corresponding to a percentage increase of 17\% to 27\%. The interval (\ref{minmax.ci}) that has a $\geq$ 95\% chance of containing the effect of the repeal on the firearm homicide rate is $[0.6,1.7]$, corresponding to an 11\% to 35\% increase in firearm homicides, providing evidence that the repeal increased firearm homicides. 
\begin{figure}[h!]
    \caption{Age-adjusted gun homicide rates per 100,000 persons in Missouri, lower control states bordering Missouri (population-weighted averages) and upper control states bordering Missouri, 1999-2016.
    }\label{bracket.trends}
    \centering
    \vspace{.5cm}
    \includegraphics[scale=0.65]{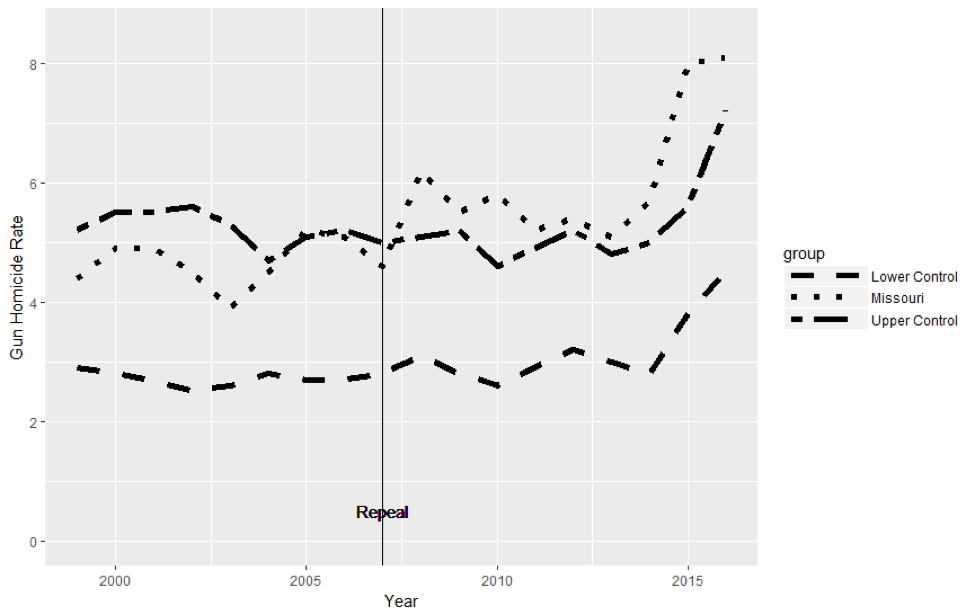}
\end{figure}
\begin{table}[h!]
\caption{Difference-in-difference estimates of effect of repeal of Missouri's permit-to-purchase handgun licensing requirement on firearm homicide rates per 100,000 persons. CI indicates confidence interval.}
\begin{center}
\begin{tabular}{|c|c|c|c|c|}
\hline Control Group & Estimate & 95\% CI & Corresponding \% Change Estimate & 95\% CI  \\ \hline
All Controls & 1.2 & [0.9, 1.5] & 24\% & [18\% ,31\%] \\ \hline
Upper Controls & 1.3 & [0.9, 1.7] & 27\% & [19\% ,35\%] \\
Lower Controls & 0.9 & [0.6, 1.2] & 17\% & [11\% ,23\%] \\
\hline
\end{tabular}
\end{center}
\label{results.table}
\end{table}
 
\subsection*{Assessing Model Assumptions: Time-Varying Confounders and Relative Trends}}
A type of time-varying confounder that is relevant to the Missouri permit-to-purchase study is a factor that only arises in the after period.  The Ferguson unrest in 2014 might have led to less effective policing
(spikes in violence typically follow social unrest) in Missouri compared to other states.  Such a time-varying confounder would be unlikely to satisfy (\ref{time.varying.confounders.assumption.i}) or (\ref{time.varying.confounders.assumption.ii}) because it arises only in the treated group (Missouri) in the after period.  However, this confounder alone does not change our finding that the repeal increased firearm homicides. If we limit the study to 2008-2013, Missouri still has larger increases in firearm homicide rates than both the upper and lower control groups; see \nameref{eAppendix6}. 

To assess the plausibility of our model (\ref{modeling.statement})-(\ref{epsilon.invariance}) and assumptions (\ref{stochastic.dominance})-(\ref{increasing.differences}), we apply the relative trends test described in \newnameref{subsec:rel.trends}. Applying the test to our study of the repeal of Missouri's permit-to-purchase law, we do not find evidence that our model assumptions are violated. Visual inspection of the relative trends of counterfactual Missouri and the upper and lower controls in the before period further supports the plausibility of our model assumptions; see \nameref{eFigure1} in \nameref{eAppendix5}.

\subsection*{Standard Error Estimates: A Poisson Model for Death Counts}
The standard errors used for inference in the previous section come directly from the CDC WONDER system. Vital statistics that derive from complete counts of deaths (by cause) are not subject to sampling error. Nonetheless, a stochastic model of vital statistics may be justified by the presence of biological, environmental, sociological, and other natural sources of variability\citep{brillinger1986biometrics}. For inferential purposes, a census may be viewed as a realization from such a stochastic process under similar conditions to those observed\citep{keyfitz1966sampling}. In particular, the observed firearm homicide death rate in any state-year may be viewed as one of a large series of possible Poisson distributed outcomes under similar conditions\citep{us2004vital}. The standard errors reported by the CDC are computed under this Poisson model.


\subsection*{A Placebo Study: Assessing Alternative Sources of Uncertainty} 
There may be other sources of uncertainty unaccounted for by the natural variability of a Poisson model for yearly state-level firearm homicides. Several recent papers suggest that such sources of uncertainty, if ignored, may yield substantially different inferential conclusions. Serially correlated data\citep{bertrand2004much}, yearly state-level shocks\citep{donald2007inference}, and small numbers of policy changes\citep{conley2011inference} can cause the standard errors returned by a fixed effects regression to be downwardly biased. We conduct a placebo study \citep{abadie2010synthetic,bertrand2004much} to address inferential challenges that arise from the presence of possibly dependent, yearly state-level shocks to the conditions that generate these Poisson realizations.

Akin to permutation inference, a placebo study in the context of the Missouri permit-to-purchase repeal analysis applies the bracketing method to every state to create a placebo intervention effect distribution. Specifically, for each state where there was no permit-to-purchase repeal we construct lower and upper control groups of neighboring states, when available, in exactly the same way we did so for Missouri. We then compute the difference-in-difference estimates using both control groups for a placebo ``repeal" on August 28, 2007. This results in two exact distributions for the placebo intervention effect estimate, one estimated using lower controls and the other using upper controls. If the permit-to-purchase repeal effect in Missouri is not spurious, we would expect to see few placebo effects greater than the ones reported in our study using either control condition.
\begin{figure}[h!]
    \caption{Histograms of placebo ``repeal" effects using different control states. (Left Panel): Histogram of placebo difference-in-difference estimates using lower control states ($n=38$ states with lower control neighbors -- includes Missouri). Two states (Oklahoma and Delaware) had a larger estimate than Missouri (dashed line).(Right Panel): Histogram of placebo difference-in-difference estimates using upper control states ($n=37$ states with upper control neighbors -- includes Missouri). One state (Delaware) had a larger estimate than Missouri (dashed line).}\label{placebo.hist}
    \centering
    \begin{tabular}{cc}
        \includegraphics[scale=0.375]{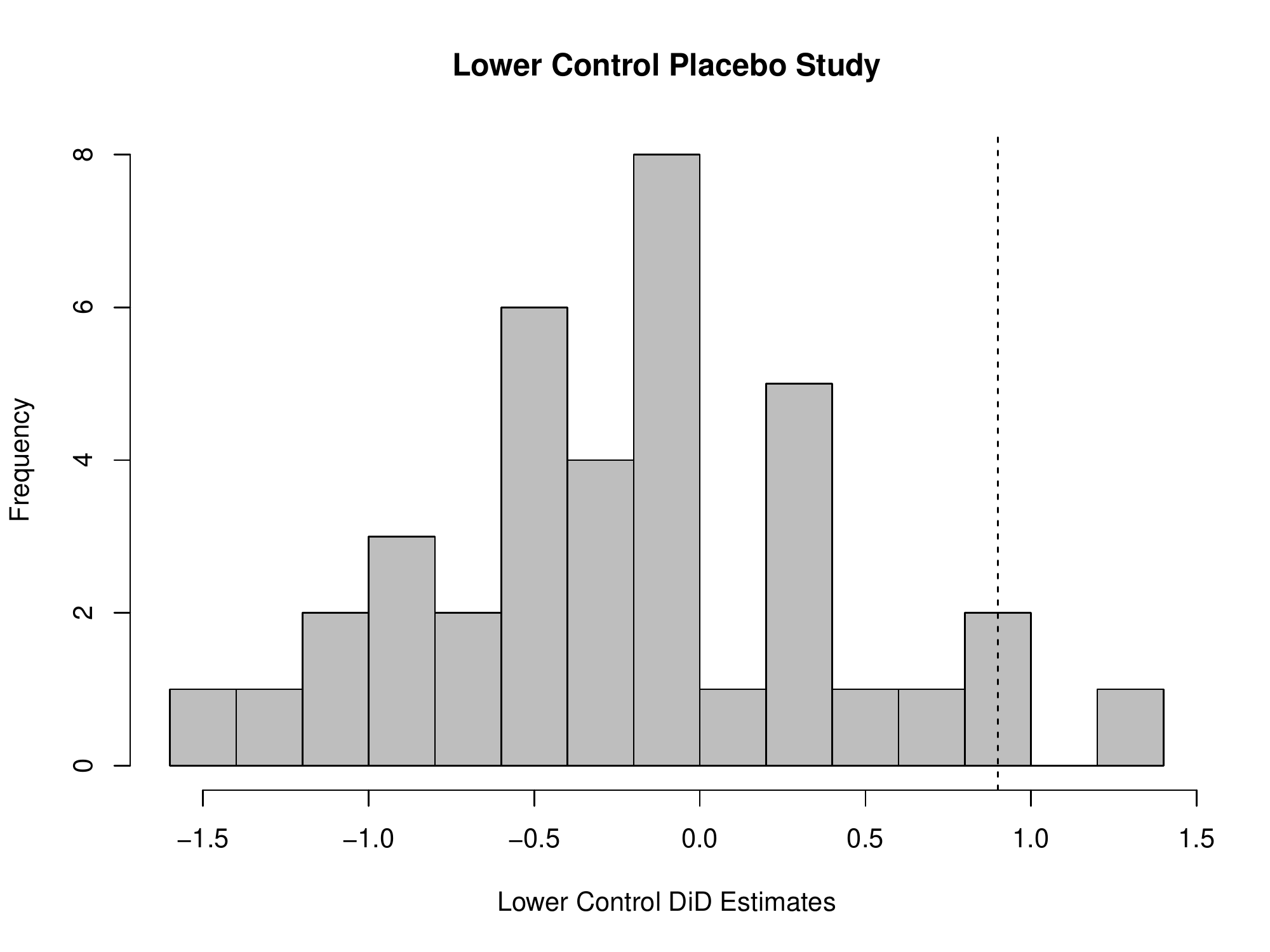} &
        \includegraphics[scale=0.375]{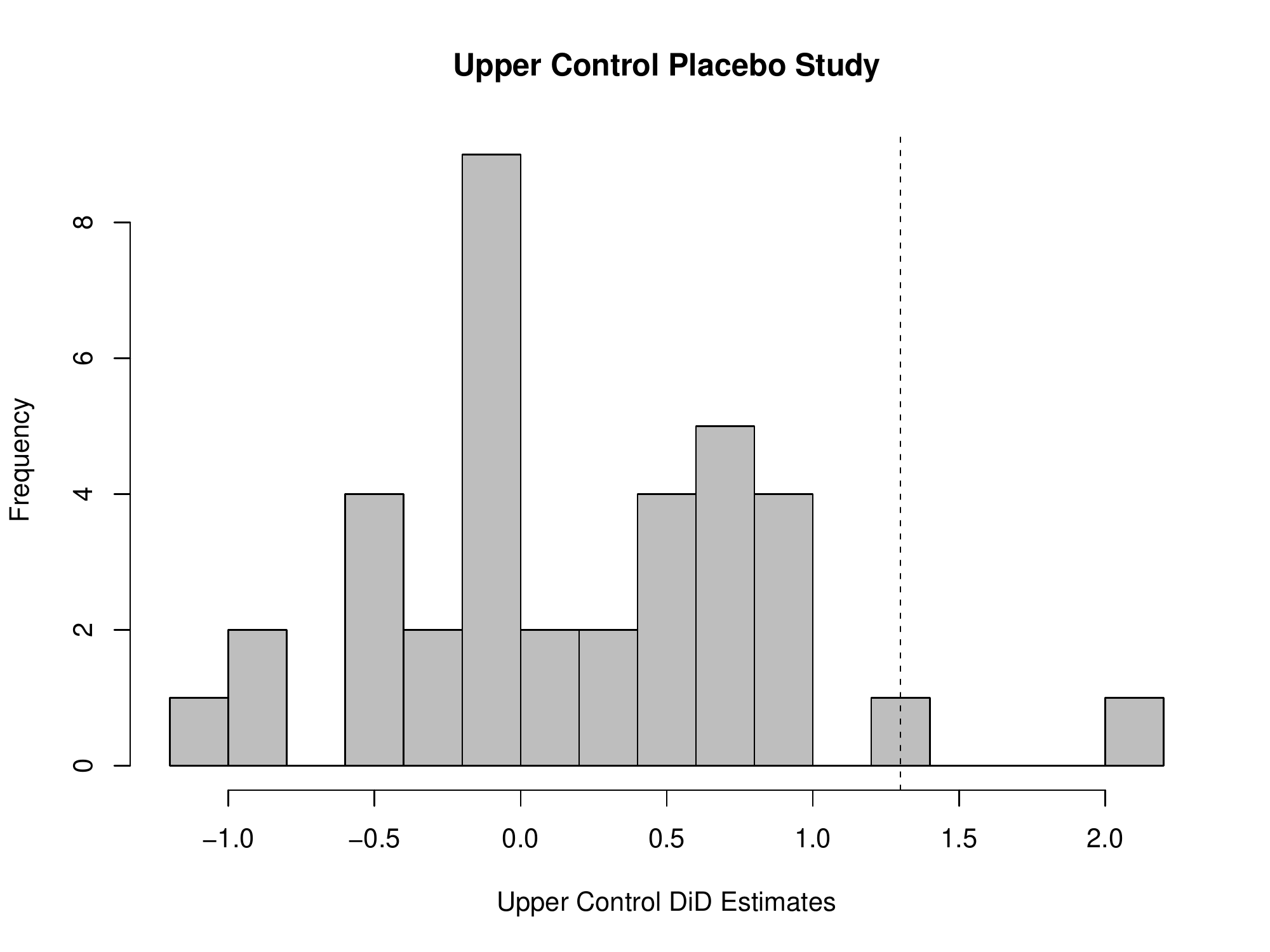}
    \end{tabular}
\end{figure}
The histograms of the placebo effects in Figure \ref{placebo.hist} suggest that the Missouri bracketing study is relatively robust to these alternative sources of variability. Of the 38 states that had lower control neighbors, only two (Oklahoma and Delaware) had placebo effect estimates using lower controls that were larger than Missouri (dashed line, left panel). Of the 37 states that had upper control neighbors, only one (Delaware) had a placebo effect estimate using upper controls that was larger than Missouri (dashed line, right panel). Alaska, Hawaii, the District of Columbia and three states with missing data in either the pre-study, before or after period were excluded from the analysis.



\section*{Conclusion and Discussion}
We developed a bracketing method for comparative interrupted time series to account for concerns that history may interact with groups.  In a study of the repeal of Missouri's permit-to-purchase handgun law, the method addressed a concern that on average, control states started out with lower firearm homicide rates than Missouri before the repeal.  Comparing both to states that started with higher firearm homicide rates than Missouri and states that started with lower rates, the repeal was associated with a significant increase in firearm homicides, thus strengthening the evidence that the repeal had a causal effect of increasing firearm homicides. 

A limitation of our estimated impact of the repeal of Missouri's permit-to-purchase law is that a Stand Your Ground  law was simultaneously adopted in Missouri.  However, in the original study by \citet{webster2014effects}, the inclusion of a Stand Your Ground indicator in the regression did not dramatically change the estimated effect. Additionally, a recent comparative interrupted time series study examining firearm homicide rates in large urban counties found that permit-to-purchase laws were associated with significant reductions in firearm homicides after controlling for the effects of Stand Your Ground laws\cite{crifasi2018association}. Further evidence that the contemporaneous Stand Your Ground law does not change the qualitative conclusion of our study can be found in the placebo study. There were 16 additional states that adopted Stand Your Ground laws within a few years of Missouri's permit-to-purchase repeal\cite{crifasi2018association}. Only one state (Oklahoma) of the 16 had a difference-in-difference placebo effect estimate using lower controls that was larger than Missouri and none of the states had placebo effect estimates using upper controls that were larger than Missouri.

Although only one of many potential patterns of bias, the history-by-group interaction bias addressed in this paper has been mentioned in the literature since at least the middle of the 20th century. A version of it is referred to {\it selection-maturation interaction} in a taxonomy of possible threats to the validity of experimental and quasi-experimental designs presented in Campbell and Stanley\citep{campbell1963experimental}. Fundamentally, bracketing relies on constructing control groups across which this potential source of confounding is systematically varied\citep{hasegawa2017}. Other methods for constructing adequate control groups in the presence of history-by-group interactions, such as the synthetic control method \citep{abadie2010synthetic}, have also found success in comparative case studies of the effect of permit-to-purchase laws on firearm homicide rates\citep{rudolph2015association}. While we do not argue that bracketing is uniformly superior to the synthetic control method, the practitioner may find that each has strengths that lend themselves to different settings. When the researcher believes that unmeasured history-by-group confounding, $h({\bf U},p)$, can be expressed as a linear factor model with time-varying slopes and group-specific loadings, the synthetic control method provides an asymptotically unbiased point estimate of the causal effect of treatment while bracketing can only provide bounds on the treatment effect. However, when the practitioner suspects that only the weaker assumptions of the model outlined in \newnameref{subsec:notation.model} hold, the bracketing bounds will remain unbiased, in that they contain the true effect in expectation, while the point estimate using synthetic controls need not be unbiased; see \nameref{eAppendix7} for further discussion. A detailed example of such a case can be found in the \nameref{eAppendix8}.

\section*{Appendix}

\subsection*{Appendix 1: Inferences Under Different Sampling Assumptions}
\labelname{Appendix 1}\label{eAppendix1} 

The standard difference-in-difference estimator using a control group $c$, $\hat{\beta}_{dd.c}$, is
\begin{eqnarray*}
\hat{\beta}_{dd.c} & = & \{ \hat{E}[Y_1|G=t,S_1=1]-\hat{E}[Y_0|G=t,S_0=1]\} \nonumber \\
& & - \{ \hat{E}[Y_1|G=c,S_1=1]-\hat{E}[Y_0|G=c,S_0=1]\}.
\end{eqnarray*}
When the samples of (i) $Y_1|G=t,S_1=1$, (ii) $Y_0|G=t,S_0=1$, (iii) $Y_1|G=c,S_1=1$ and (iv) $Y_0|G=c,S_0=1$ are independent, then the standard error of $\hat{\beta}_{dd.c}$ is
\begin{gather} 
SE(\hat{\beta}_{dd.c})= \nonumber \\
\{ SE(\hat{E}[Y_1|G=t,S_1=1])^2+SE(\hat{E}[Y_0|G=t,S_0=1])^2 \nonumber \\
+SE(\hat{E}[Y_1|G=c,S_1=1])^2+SE(\hat{E}[Y_0|G=c,S_0=1])^2 \}^{1/2}. \label{se.diff.in.diff}
\end{gather} 
We use (\ref{se.diff.in.diff}) to make inferences for our study of the effect of the repeal of Missouri's permit-to-purchase law, where the $\hat{E}$ and corresponding SEs are obtained from the CDC's WONDER system.

Let $\kappa_{tt}$ be the \% change in the treated group's mean outcome in the after period compared to its mean counterfactual outcomes in the after period in the absence of treatment,
\[
\kappa_{tt}=100\times \frac{E[Y_1^{(1)}|G=t,S_1=1]-E[Y_1^{(0)}|G=t,S_1=1]}{E[Y_1^{(0)}|G=t,S_1=1]}.
\]
An estimate of $\kappa_{tt}$ using control group $c$ and assuming the parallel trends of standard-in-differences is
\[
\hat{\kappa}_{tt.c}= 100\times \frac{\hat{\beta}_{dd.c}}{\hat{E}(Y_0|G=t,S_0=1)+\{ \hat{E}(Y_1|G=c,S_1=1)-\hat{E}(Y_0|G=c,S_0=1)\}}.
\]
We approximate the standard error of $\hat{\kappa}_{tt.c}$ using the Delta method.

The model (\ref{modeling.statement}) can be extended to allow for observed covariates, clustering and multiple time points using a regression framework \citep{imbens2009recent}.  The difference-in-difference estimator may be computed by regressing the observed outcome $Y$ on a time period dummy, a group dummy and a treatment variable.  Observed covariates ${\bf{X}}_{ip}$ that could vary by time can be incorporated into the model and then the difference-in-difference regression estimator can be computed by regressing $Y$ on the observed covariates, a time period dummy, a group dummy and a treatment variable.  The model assumptions then need to hold only conditionally on the observed covariates.  The comparative interrupted time series can be applied to settings with more than two time periods.  A full set of time period dummies can be added to model (\ref{modeling.statement}).  The effect of the treatment over time can be allowed to vary by interacting the treatment dummy with time.

Within each group, there may be clusters of units, e.g., different countries that had the same policy reform.  For such settings, we can extend model (\ref{modeling.statement}) to the following \citep{donald2007inference} where the index $cip$ denotes the $i$th unit in cluster $c$ at time period $p$:
\begin{equation}
Y_{cip}^{(d)}=h({\bf{U}}_{cip},p)+\beta d + \eta_{cp}+\epsilon_{cip}, \label{clustering.model}
\end{equation}
where $\eta_{cp}$ represents an effect shared by members of cluster $c$ in period $p$, e.g., an economic shock that is specific to a country $c$ in period $p$.  Under an assumption that the $\eta_{cp}$ are independent and identically distributed (i.i.d.) normal random variables,  \citet{donald2007inference} showed that if we compute the mean in each cluster at each time period, and regress these cluster/period means on fixed effects for each cluster, a time period dummy and a treatment variable, then the $t$ statistic for the treatment variable ($\frac{\hat{\beta}-\beta}{SE(\hat{\beta})}$) has a $t$ distribution with the number of clusters minus two degrees of freedom.  Using this approach, we do not need to have individual data but only summary data for each cluster.  Other approaches to inference that allow for the $\eta_{cp}$ to be non-i.i.d. such as autocorrelated within group, have been developed. \citep{bertrand2004much,hansen2007generalized}.

Note that the presence of at least two clusters in at least one group enables us to make inferences that allow for shared effects $\eta_{cp}$.  When there is only one cluster in each group, e.g., we are comparing just two countries, one in which a policy reform was implemented and one in which it was not, then there are zero degrees of freedom to estimate the variance of the $\eta_{cp}$ so inferences cannot be drawn that allow for $\eta_{cp}$ to be nonzero using data from entirely within the sample.  For such settings, it may be possible to get information from outside the sample to get a plausible estimate of the variance of the $\eta_{cp}$ \citep{blitstein2005increasing,donald2007inference}.

\subsection*{Appendix 2: Proof of (\ref{bracketing.result}) in \newnameref{subsec:bracketing}}
\labelname{Appendix 2}\label{eAppendix2}  

Suppose $h({\bf{U}},1)-h({\bf{U}},0)$ is a bounded increasing function of ${\bf{U}}$.  Then from (\ref{stochastic.dominance}) and the property that bounded increasing functions of stochastically ordered random variables preserve order, it follows that
\begin{equation}
E[\hat{\beta}_{dd.uc}]\leq \beta \leq E[\hat{\beta}_{dd.lc}]. \label{bracketing1}
\end{equation}
Similarly, if $h({\bf{U}},1)-h({\bf{U}},0)$ is a bounded decreasing function of ${\bf{U}}$,
\begin{equation}
E[\hat{\beta}_{dd.lc}]\leq \beta \leq E[\hat{\beta}_{dd.uc}]. \label{bracketing2}
\end{equation}
(\ref{bracketing.result}) follows from (\ref{bracketing1}) and (\ref{bracketing2}).

\subsection*{Appendix 3: Proof for Result in \newnameref{subsec:inference}}
\labelname{Appendix 3}\label{eAppendix3}

Here we prove that (\ref{minmax.ci}) has probability $\geq 1-\alpha$ of containing both $\min (\theta_{lc.t},\theta_{uc.t})$ and $\max (\theta_{lc.t},\theta_{uc.t})$ under the assumption that the two sided CIs are constructed in the usual way by taking the union of two one-sided $1-(\alpha /2)$ confidence intervals.  The result is basically derived by inverting multiparameter hypothesis tests about the minimum or maximum of two parameters \citep{lehmann1952testing,berger1982multiparameter}.  Let $q=\min (\theta_{lc.t},\theta_{uc.t})$ and $r=\max (\theta_{lc.t},\theta_{uc.t})$.  The probability that (\ref{minmax.ci}) does not contain both $\min (\theta_{lc.t},\theta_{uc.t})$ and $\max (\theta_{lc.t},\theta_{uc.t})$ is bounded by the probability that $q$ is less than the lower endpoint of the interval plus the probability that $r$ is greater than the upper endpoint of the interval.  The probability that $q$ is less than the lower endpoint of the interval is the probability that both one-sided tests
$H_0^l: \theta_{lc.t}\leq q$ vs. $H_1^l: \theta_{lc.t}>q$ and $H_0^u: \theta_{uc.t}\leq q$ vs. $H_1^l: \theta_{uc.t}>q$ give p-values $\leq \alpha /2$, which has probability at most $\alpha /2$ since each individual event has probability at most $\alpha /2$.  Similarly, the probability that $r$ is greater than the upper endpoint of the interval is the probability that both one-sided tests
$H_0^{l'}: \theta_{lc.t}\geq r$ vs. $H_1^{l'}: \theta_{lc.t}<r$ and $H_0^{u'}: \theta_{uc.t}\geq r$ vs. $H_1^{l'}: \theta_{uc.t}<r$ give p-values $\leq \alpha /2$, which has probability at most $\alpha /2$ since each individual event has probability at most $\alpha /2$.  Thus, the probability that (\ref{minmax.ci}) does not contain both $\min (\theta_{lc.t},\theta_{uc.t})$ and $\max (\theta_{lc.t},\theta_{uc.t})$ is bounded by $\alpha$.

\subsection*{Appendix 4: Modeling Time-varying Confounders}
\labelname{Appendix 4}\label{eAppendix4}

We model a setting with time-varying confounders as follows.  We maintain the assumptions in Section \newnameref{subsec:notation.model} except for (\ref{epsilon.invariance}).
We let ${\bf{U}}$ contain all variables that affect the outcome that differ in distribution between the groups (treated, upper control, lower control) in the before period and let $\epsilon_0$ summarize the effect of factors in the before period that do not differ in distribution between the groups.  We can model the average effect of the factors in $\epsilon_0$ as an intercept in the $h({\bf{U}},0)$ function so that $E(\epsilon_0|S_0=1,G=g)=0$ holds for all groups $g=lc,uc,tc$.  The effect of factors that do not differ in distribution between the groups in the after period as well as the effect of time-varying confounders in the after period are summarized in $\epsilon_1$.  Some of these time-varying confounders may be variables in ${\bf{U}}$ that have changed their level over time.  Let ${\bf{U}}_0\equiv {\bf{U}}$ be the value of the variables in ${\bf{U}}$ in the before period and ${\bf{U}}_1$ be their value in the after period, where ${\bf{U}}_0={\bf{U}}_1$ for a unit only in the population in the after period (with ${\bf{U}}$ defined this way, the validity of (\ref{time.invariance}) needs to be considered carefully).  Then, assuming that the average effect of the factors in $\epsilon_1$ that do not differ between the groups in the after period is modeled as an intercept in $h({\bf{U}},1)$, we have
\[
E(\epsilon_1|G=g,S_1=1)=E[h({\bf{U}}_1,1)-h({\bf{U}}_0,1)|G=g,S_1=1].
\]
Then for (\ref{time.varying.confounders.assumption.i}) to hold, we need to have
\begin{gather}
E[h({\bf{U}}_1,1)-h({\bf{U}}_0,1)|G=uc,S_1=1]\geq E[h({\bf{U}}_1,1)-h({\bf{U}}_0,1)|G=t,S_1=1] \nonumber \\
\geq E[h({\bf{U}}_1,1)-h({\bf{U}}_0,1)|G=lc,S_1=1] \label{equivalent.condition.time.varying.confounders.assumption.i}
\end{gather}

A set of sufficient conditions for (\ref{equivalent.condition.time.varying.confounders.assumption.i}) to hold when ${\bf{U}}$ is univariate and the assumptions in Section \newnameref{subsec:notation.model} hold is the following: (a) $S_0=S_1=1$ for all units so that all units are in the study population in both periods; (b) $U_1-U_0$ is independent of $U_0$ given $G$; (c) the function $h(U,1)$ is convex in $U$ so that $h$ has increasing differences in the sense that for $u$, $u^{'}$, $u^{''}$, $u^{'''}$ such that $u-u^{'}=u^{''}-u^{'''}$ and $u>u^{''}$, the following inequality holds: $h(u,1)-h(u^{'},1)\geq h(u^{''},1)-h(u^{'''},1)$, and (d) $U_1-U_0|G=lc \preceq U_1-U_0|G=t \preceq U_1-U_0|G=uc$.  The proof that this set of sufficient conditions implies that (\ref{equivalent.condition.time.varying.confounders.assumption.i}) holds is as follows.  Let $D_{lc}$ be a random variable with the distribution of $U_1-U_0|G=lc$ where $D_{lc}$ is independent of $U_0$ given $G$.  Then from (c) and (\ref{stochastic.dominance}), it follows that
\begin{equation}
E[h(U_0+D_{lc},1)-h(U_0,1)|G=t]\geq E[h(U_0+D_{lc},1)-h(U_0,1)|G=lc]. \label{implication.from.c.stochdom}
\end{equation}
Now let $D_{t}$ be a random variable with the conditional distribution of $U_1-U_0|G=t$ and $D_{uc}$ be a random variable with the conditional distribution of $U_1-U_0|G=uc$ where $D_t$ and $D_{uc}$ are independent of $U_0$ given $G$.  Then from (d) and $h$ being an increasing function, it follows that $E[h(U_0+D_{t})|G=t]\geq E[h(U_0+D_{lc})|G=t]$.  Combining this with (\ref{implication.from.c.stochdom}), we have
\[
E[h(U_0+D_{t},1)-h(U_0,1)|G=t]\geq E[h(U_0+D_{lc},1)-h(U_0,1)|G=lc]
\]
which is equivalent to
\begin{equation}
E[h(U_1,1)-h(U_0,1)|G=t]\geq E[h(U_1,1)-h(U_0,1)|G=lc]. \label{equivalence.t.lc}
\end{equation}
Similarly from (d) and $h$ being an increasing function, it follows that $E[h(U_0+D_{uc})|G=uc]\geq E[h(U_0+D_t)|G=uc]$, and from (c) and (\ref{stochastic.dominance}), it follows that
\[
E[h(U_0+D_{t},1)-h(U_0,1)|G=uc] \geq E[h(U_0+D_{t},1)-h(U_0,1)|G=t],
\]
and combining these, we have that
\[
E[h(U_0+D_{uc},1)-h(U_0,1)|G=uc]\geq E[h(U_0+D_{t},1)-h(U_0,1)|G=t]
\]
which is equivalent to
\begin{equation}
E[h(U_1,1)-h(U_0,1)|G=uc]\geq E[h(U_1,1)-h(U_0,1)|G=t]. \label{equivalence.t.uc}
\end{equation}
Combining (\ref{equivalence.t.lc}) and (\ref{equivalence.t.uc}) gives us the desired conclusion.

{\it{Proof that (\ref{bracketing.result}) still holds as long as when (i) in (\ref{increasing.differences}) holds, (\ref{time.varying.confounders.assumption.i}) holds or when (ii) in (\ref{increasing.differences}) holds, (\ref{time.varying.confounders.assumption.ii}) holds}}.  When there are time varying confounders, we have that $E[\hat{\beta}_{dd.lc}]$ is the expression on the right hand side of (\ref{diff.in.diff.bias.lc}) plus $E(\epsilon_1|G=t,S_1=1)-E(\epsilon_1|G=lc,S_0=1)$ and $E[\hat{\beta}_{dd.lc}]$ is the expression on the right hand side of (\ref{diff.in.diff.bias.uc}) plus $E(\epsilon_1|G=t,S_1=1)-E(\epsilon_1|G=uc,S_0=1)$.  When (i) in (\ref{increasing.differences}) holds, the expression on the right hand side of (\ref{diff.in.diff.bias.lc}) is $\geq \beta$ and the expression on the right hand side of (\ref{diff.in.diff.bias.uc}) is $\leq \beta$.  Combining the facts in the last two sentences, we have that if (i) in (\ref{increasing.differences}) and (\ref{time.varying.confounders.assumption.i}) holds, $E[\hat{\beta}_{dd.uc}]\leq \beta \leq E[\hat{\beta}_{dd.lc}]$ and if (ii) in (\ref{increasing.differences}) and (\ref{time.varying.confounders.assumption.ii}) holds, $E[\hat{\beta}_{dd.lc}]\leq \beta \leq E[\hat{\beta}_{dd.uc}]$.

\subsection*{Appendix 5: Test of Model/Assumptions by Examining the Groups' Relative Trends in the Before Period}
\labelname{Appendix 5}\label{eAppendix5}

We can test whether the violating pattern (iii) is present in the before period using an intersection-union test \citep{lehmann1952testing,berger1982multiparameter}, which find evidence (say p-value $<0.05$) for (iii) if there is evidence (p-value $<.05$) for both (a) the difference between the upper control group and the counterfactual treated group is larger in the second part of the before period than the first part and (b) the difference between the counterfactual treated group and the lower control group is smaller in the second part than the first part; for the firearm homicide data, splitting the before period into the two parts, 1999-2002 and 2003-2007, (a) gives a p-value of $0.96$ and (b) gives a p-value of $0.5$, so there is not evidence for (iii) being violated.  Pattern (iv) can be tested in a similar way and for the firearm homicide data, there is not evidence for pattern (iv) holding (p-values of $0.04$ and $0.5$). Ideally, this testing procedure should have sufficient power to reduce the chance of proceeding with the analysis when the assumptions of the model don't, in fact, hold to an acceptable level. When sample sizes are beyond the control of the investigator or, for example, when dealing with complete counts of firearm homicides where variability depends on the rate itself rather than sampling error, increasing the level of the test can achieve some improvement in power. The p-value is $\ge 0.5$ for the test of each alternative, that (iii) holds and that (iv) holds. Hence, $\alpha$ would have to be increased beyond $0.5$ to affect the conclusions about the plausibility of our model assumptions.

\begin{figure}[h!]
    \caption{(Left Panel): Relative trends of the population-weighted firearm homicide rates for the upper (dashed blue) and lower (dashed red) groups and the counterfactual treated group (dashed black) over the before period. The vertical bars indicate 95\% CIs. (Right Panel): Relative trends of the population-weighted firearm homicide rates for the upper (dashed blue) and lower (dashed red) groups and the treated group (dashed black) over the entire period. The vertical bars indicate 95\% CIs.  }\labelname{eFigure 1}\label{eFigure1}
    \centering
    \begin{tabular}{cc}
        \includegraphics[scale=0.6]{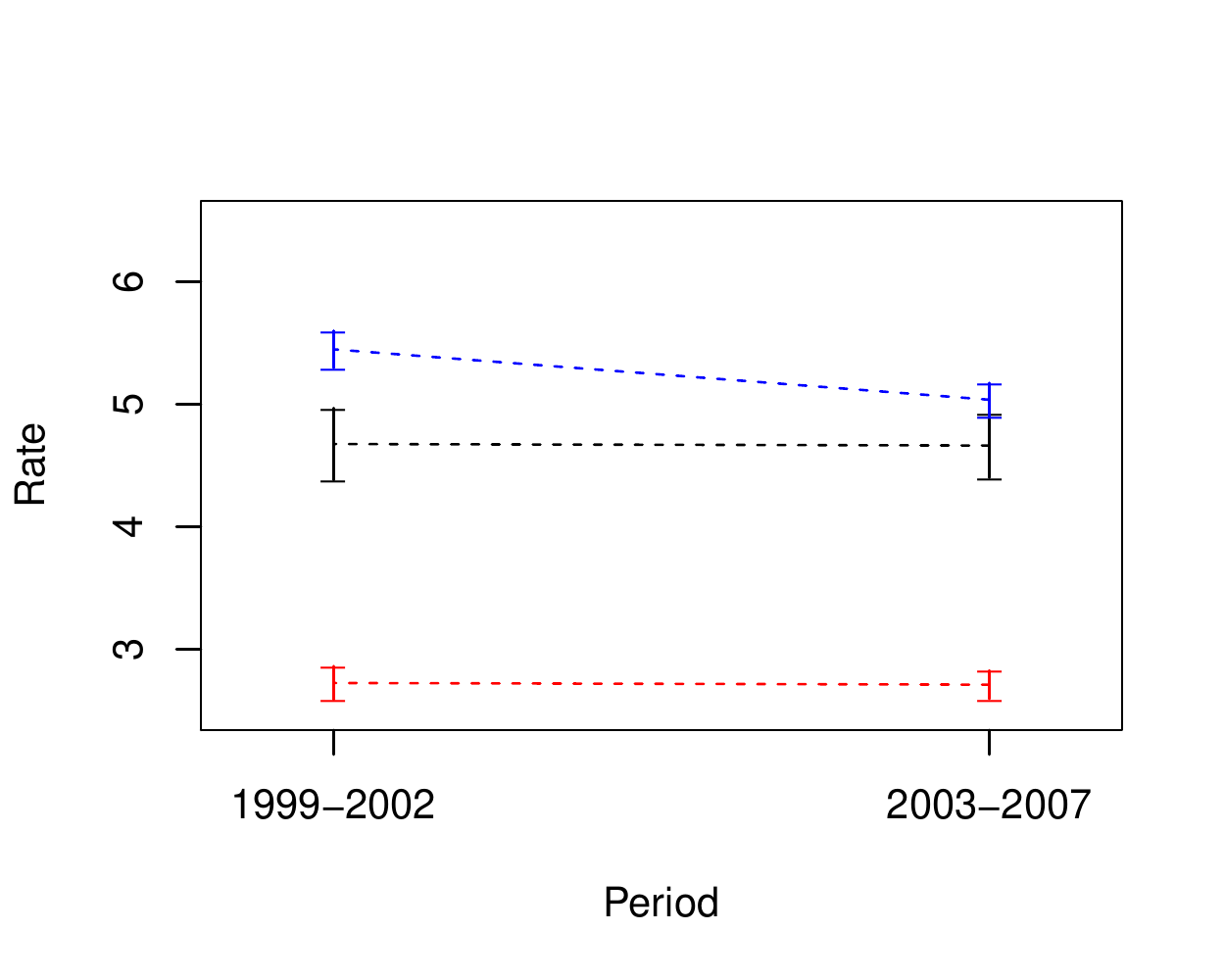} & 
        \includegraphics[scale=0.6]{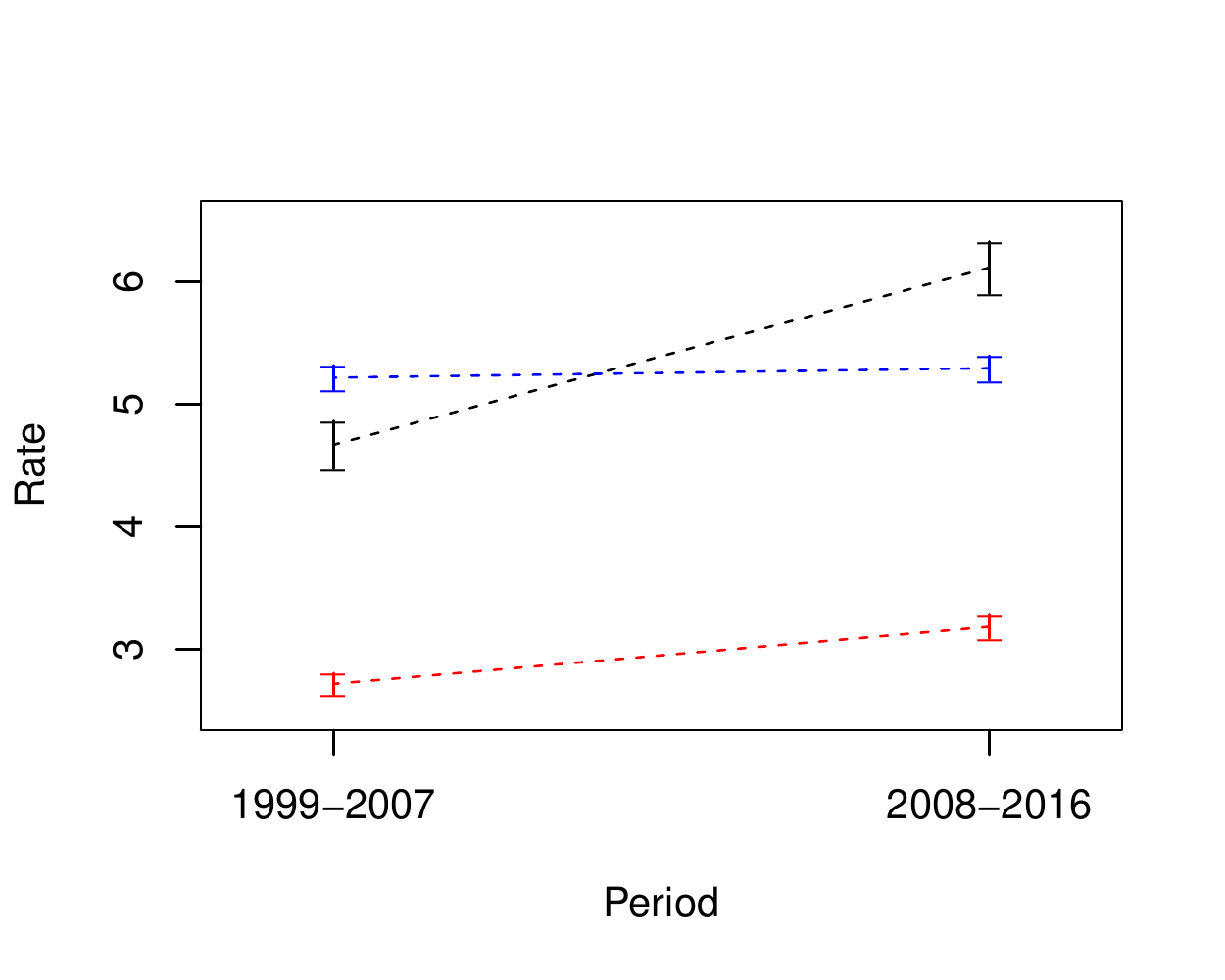}
    \end{tabular}
\end{figure}

Alternatively, the presence of violating patterns (iii) and (iv) can be assessed visually without requiring a formal testing procedure. In the left panel of \nameref{eFigure1} we plot the relative trends of the population-weighted firearm homicide rates for the upper (dashed blue) and lower (dashed red) groups and the counterfactual treated group (dashed black) over the before period. The vertical bars indicate 95\% CIs. Visually, there is no strong evidence that pattern (iii) or (iv) is present. The difference between upper controls and counterfactual Missouri and between counterfactual Missouri and the lower controls both get smaller in the latter part of the before period. We can also partially assess whether this pattern might hold over the entire study period, our primary concern, by addressing how the upper and lower control trends compare between the before period and the entire study period. In the right panel of \nameref{eFigure1} we plot the relative trends of the two control groups and treated group over the entire study period. The dashed black lines are not comparable between panels because the left panel is a counterfactual trend whereas the trend in the right panel is subject to treatment (i.e. permit-to-purchase repeal). However, we can assess the comparability of the pattern of the control group trends between the two panels. They appear similar, with a slight narrowing of the difference in population-weighted firearm homicide rates over time. 

When paired with the test described above, visual inspection can answer questions about our model assumptions that our intersection-union tests do not address directly: If we find evidence that pattern (iii) or (iv) is present, are the violations substantial enough to arrest the planned analysis or should we still proceed but with increased caution? If the test doesn't find evidence of a violation is that because our assumptions hold, at least approximately, or is it due to large standard errors and/or low power? We recommend that testing and visual inspection should be used in conjunction when assessing the plausibility of the model assumptions.

If one does find evidence for pattern (iii) or (iv) holding in the before period, and if one thinks there has been a structural shift such that the model (\ref{modeling.statement})-(\ref{epsilon.invariance}) and assumptions (\ref{stochastic.dominance})-(\ref{increasing.differences}) only start to hold in the latter part of the before period but continue to hold in the after period, one could just use the latter part of the before period.  This is similar to the scenario in a difference-in-difference model when there is evidence of a diverging trend during an earlier portion of the pre-intervention period, researchers can restrict the analysis to include only the latter part of the before period with the hope that parallel trend assumption is more likely to be valid\citep{volpp2007mortality}.  However, the finding of pattern (iii) or (iv) in the before period suggests caution.

\subsection*{Appendix 6: Analysis Using After Period of 2008-2013}
\labelname{Appendix 6}\label{eAppendix6}

For the period of 2008-2013, Missouri's age-adjusted firearm homicide rate was 5.5, the upper control group's age-adjusted firearm homicide rate was 5.0 and the lower control's age adjusted firearm homicide rate was 2.9.  Using an after period of 2008-2013, difference-in-difference estimates for the upper and lower control groups are shown in Table \ref{results.table.2008.2013}.  Using an after period of 2008-2013, the interval (\ref{minmax.ci}) that has a $\geq$ 95\% chance of containing the effect of the repeal on the firearm homicide rate is $[0.2,1.4]$, corresponding to a 5\% to 31\% increase in firearm homicides, providing evidence that the repeal increased firearm homicides.

\begin{table}[h!]
\caption{Difference-in-difference estimates of effect of repeal of Missouri's permit-to-purchase handgun licensing requirement on firearm homicide rates per 100,000 persons using after period of 2008-2013}
\begin{center}
\begin{tabular}{|c|c|c|c|c|}
\hline Control Group & Estimate & 95\% CI & Corresponding \% Change Estimate & 95\% CI  \\ \hline
Upper Controls & 1.0 & [0.6, 1.4] & 22\% & [14\% ,31\%] \\
Lower Controls & 0.6 & [0.2, 1.0] & 17\% & [5\% ,19\%] \\
\hline
\end{tabular}
\end{center}
\label{results.table.2008.2013}
\end{table}

\subsection*{Appendix 7: Comparison with the Synthetic Control Method}
\labelname{Appendix 7}\label{eAppendix7}

\citet{abadie2010synthetic} proposed constructing a synthetic control group which is a linear combination of multiple control groups that matches the before period outcomes of the treatment group.
The synthetic control method provides asymptotically unbiased estimates of the causal effect of treatment assuming that the unmeasured confounders can be represented by a factor model with the factors' effects in each time period being linear with a time-specific slope, whereas our bracketing method only provides bounds under this assumption. However, this assumption is strong and is not generally satisfied in our model (\ref{modeling.statement})-(\ref{epsilon.invariance}). In the following section we provide a simple example that satisfies the assumptions of our model but for which the estimate returned by the synthetic control method will be biased.

If the types of interaction between history and group in the after period that are of concern have occurred in the before period (e.g., a similar recession occurred in the after period as the before period), then the synthetic control method's matching of the before period outcomes might enable it to match the treated group's counterfactual trajectory in the after period in the absence of treatment.  However, if the types of interaction are different (e.g., there is a more severe recession in the after period or the interactions between poor health and the macroeconomy have been altered by other policy changes), then the synthetic control's matching in the before period does not provide much reassurance unless one has a basis for strong functional form assumptions such as the factors representing the unmeasured confounders' having a linear effect in each time period.  In contrast, the bracketing method relies on assumptions such as (\ref{increasing.differences}) that the unmeasured confounders' effect is increasing (or decreasing) in importance over time over the whole range of the unmeasured confounders that can be assessed using subject matter knowledge without making strong functional form assumptions.

\subsection*{Appendix 8: Example of How Synthetic Control Model Assumptions Are Violated in Our Model}
\labelname{Appendix 8}\label{eAppendix8}

For example, suppose $U$ has an exponential distribution in each group with scale 0.2, 0.5 and $\tau$ in the lower control, upper control and treated groups respectively where $0.2<\tau <0.5$ and $h(U,0)=U$, $h(U,1)=exp(U)$.  Then the synthetic control linear combination is $\frac{\tau -0.2}{0.3}\times$ lower control group + $\frac{0.5-\tau}{0.3}\times$ upper control group.  For the after period, the linear combination of the mean outcomes for the synthetic control linear combination is $\frac{\tau -0.2}{0.3}\times 1.25 + \frac{0.5-\tau}{0.3}\times 2$ while the treated group's counterfactual mean outcome in the absence of treatment is
$\frac{-1/\tau}{-1/\tau + 1}$, and $\frac{\tau -0.2}{0.3}\times 1.25 + \frac{0.5-\tau}{0.3}\times 2 < \frac{-1/\tau}{-1/\tau + 1}$ for all $0.2<\tau <0.5$.  Thus the synthetic control group's after period mean is always less than than the counterfactual after period mean for the treatment group in the absence of treatment.

 \bibliographystyle{IEEEtranN}

\bibliography{dylan_refs_v1}

\begin{thebibliography}{34}
\providecommand{\natexlab}[1]{#1}
\providecommand{\url}[1]{#1}
\csname url@samestyle\endcsname
\providecommand{\newblock}{\relax}
\providecommand{\bibinfo}[2]{#2}
\providecommand{\BIBentrySTDinterwordspacing}{\spaceskip=0pt\relax}
\providecommand{\BIBentryALTinterwordstretchfactor}{4}
\providecommand{\BIBentryALTinterwordspacing}{\spaceskip=\fontdimen2\font plus
\BIBentryALTinterwordstretchfactor\fontdimen3\font minus
  \fontdimen4\font\relax}
\providecommand{\BIBforeignlanguage}[2]{{%
\expandafter\ifx\csname l@#1\endcsname\relax
\typeout{** WARNING: IEEEtranN.bst: No hyphenation pattern has been}%
\typeout{** loaded for the language `#1'. Using the pattern for}%
\typeout{** the default language instead.}%
\else
\language=\csname l@#1\endcsname
\fi
#2}}
\providecommand{\BIBdecl}{\relax}
\BIBdecl

\bibitem[Cook et~al.(2002)Cook, Campbell, and Shadish]{cook2002experimental}
T.~D. Cook, D.~T. Campbell, and W.~Shadish, \emph{Experimental and
  quasi-experimental designs for generalized causal inference}.\hskip 1em plus
  0.5em minus 0.4em\relax Boston, MA: Houghton Mifflin, 2002.

\bibitem[Meyer(1995)]{meyer1995natural}
B.~D. Meyer, ``Natural and quasi-experiments in economics,'' \emph{Journal of
  Business \& Economic Statistics}, vol.~13, no.~2, pp. 151--161, 1995.

\bibitem[Bernal et~al.(2017)Bernal, Cummins, and
  Gasparrini]{bernal2017interrupted}
J.~L. Bernal, S.~Cummins, and A.~Gasparrini, ``Interrupted time series
  regression for the evaluation of public health interventions: a tutorial,''
  \emph{International journal of epidemiology}, vol.~46, no.~1, pp. 348--355,
  2017.

\bibitem[Wing et~al.(2018)Wing, Simon, and Bello-Gomez]{wing2018designing}
C.~Wing, K.~Simon, and R.~A. Bello-Gomez, ``Designing difference in difference
  studies: Best practices for public health policy research,'' \emph{Annual
  Review of Public Health}, vol.~39, pp. 453--469, 2018.

\bibitem[Cook and Campbell(1979)]{cook1979}
T.~D. Cook and D.~T. Campbell, \emph{Quasi-experimentation: {D}esign and
  analysis issues for field settings}.\hskip 1em plus 0.5em minus 0.4em\relax
  Chicago, IL: Rand McNally, 1979.

\bibitem[Reynolds and West(1987)]{reynolds1987multiplist}
K.~D. Reynolds and S.~G. West, ``A multiplist strategy for strengthening
  nonequivalent control group designs,'' \emph{Evaluation Review}, vol.~11,
  no.~6, pp. 691--714, 1987.

\bibitem[Campbell(1969)]{campbell1969}
D.~T. Campbell, ``Prospective: Artifact and control,'' in \emph{Artifacts in
  Behavioral Research: Robert Rosenthal and Ralph L. Rosnow's Classic Books},
  R.~Rosenthal and R.~L. Rosnow, Eds.\hskip 1em plus 0.5em minus 0.4em\relax
  New York, NY: Academic Press, 1969, pp. 351--382.

\bibitem[Rosenbaum(1987)]{rosenbaum1987}
P.~R. Rosenbaum, ``The role of a second control group in an observational
  study,'' \emph{Statistical Science}, vol.~2, no.~3, pp. 292--306, 1987.

\bibitem[Athey and Imbens(2006)]{athey2006identification}
S.~Athey and G.~W. Imbens, ``Identification and inference in nonlinear
  difference-in-differences models,'' \emph{Econometrica}, vol.~74, no.~2, pp.
  431--497, 2006.

\bibitem[Shaked and Shanthikumar(1994)]{shakedstochastic}
M.~Shaked and J.~G. Shanthikumar, \emph{Stochastic Orders and Their
  Applications}.\hskip 1em plus 0.5em minus 0.4em\relax New York: Academic
  Press, 1994.

\bibitem[Dix-Carneiro and Kovak(2017)]{dix2017trade}
R.~Dix-Carneiro and B.~K. Kovak, ``Trade liberalization and regional
  dynamics,'' \emph{American Economic Review}, vol. 107, no.~10, pp. 2908--46,
  2017.

\bibitem[Burstein and Vogel(2017)]{burstein2017international}
A.~Burstein and J.~Vogel, ``International trade, technology, and the skill
  premium,'' \emph{Journal of Political Economy}, vol. 125, no.~5, pp.
  1356--1412, 2017.

\bibitem[Volpp et~al.(2007)Volpp, Rosen, Rosenbaum, Romano, Even-Shoshan, Wang,
  Bellini, Behringer, and Silber]{volpp2007mortality}
K.~G. Volpp, A.~K. Rosen, P.~R. Rosenbaum, P.~S. Romano, O.~Even-Shoshan,
  Y.~Wang, L.~Bellini, T.~Behringer, and J.~H. Silber, ``Mortality among
  hospitalized medicare beneficiaries in the first 2 years following
  {A}{C}{G}{M}{E} resident duty hour reform,'' \emph{Journal of the American
  Medical Association}, vol. 298, no.~9, pp. 975--983, 2007.

\bibitem[Webster et~al.(2014)Webster, Crifasi, and Vernick]{webster2014effects}
D.~Webster, C.~K. Crifasi, and J.~S. Vernick, ``Effects of the repeal of
  missouri�s handgun purchaser licensing law on homicides,'' \emph{Journal of
  Urban Health}, vol.~91, no.~2, pp. 293--302, 2014.

\bibitem[CDC()]{CDC2018}
Centers for Disease Control and Prevention, National Center for Health
  Statistics. Compressed Mortality File on CDC WONDER Online Database, released
  December 2017. Data are from the Compressed Mortality File 1999-2016 Series
  20 No. 2V, 2017, as compiled from data provided by the 57 vital statistics
  jurisdictions through the Vital Statistics Cooperative Program. Accessed at
  http://wonder.cdc.gov on Mar 19, 2018.

\bibitem[Matthews et~al.(2006)Matthews, Shepherd, and
  Sivarajasingham]{matthews2006violence}
K.~Matthews, J.~Shepherd, and V.~Sivarajasingham, ``Violence-related injury and
  the price of beer in england and wales,'' \emph{Applied Economics}, vol.~38,
  no.~6, pp. 661--670, 2006.

\bibitem[Wolf et~al.(2014)Wolf, Gray, and Fazel]{wolf2014violence}
A.~Wolf, R.~Gray, and S.~Fazel, ``Violence as a public health problem: An
  ecological study of 169 countries,'' \emph{Social Science \& Medicine}, vol.
  104, pp. 220--227, 2014.

\bibitem[Shepherd and Page(2015)]{shepherd2015economic}
J.~Shepherd and N.~Page, ``The economic downturn probably reduced violence far
  more than licensing restrictions,'' \emph{Addiction}, vol. 110, no.~10, pp.
  1583--1584, 2015.

\bibitem[Brillinger(1986)]{brillinger1986biometrics}
D.~R. Brillinger, ``A biometrics invited paper with discussion: the natural
  variability of vital rates and associated statistics,'' \emph{Biometrics},
  vol.~42, pp. 693--734, 1986.

\bibitem[Keyfitz(1966)]{keyfitz1966sampling}
N.~Keyfitz, ``Sampling variance of standardized mortality rates,'' \emph{Human
  Biology}, vol.~38, no.~3, pp. 309--317, 1966.

\bibitem[{US Department of Health and Human Services}
  et~al.(2004)]{us2004vital}
{US Department of Health and Human Services} \emph{et~al.}, ``Vital statistics
  of the united states: mortality, 1999 technical appendix,'' \emph{National
  Center for Health Statistics}, 2004.

\bibitem[Bertrand et~al.(2004)Bertrand, Duflo, and
  Mullainathan]{bertrand2004much}
M.~Bertrand, E.~Duflo, and S.~Mullainathan, ``How much should we trust
  differences-in-differences estimates?'' \emph{The Quarterly Journal of
  Economics}, vol. 119, no.~1, pp. 249--275, 2004.

\bibitem[Donald and Lang(2007)]{donald2007inference}
S.~G. Donald and K.~Lang, ``Inference with difference-in-differences and other
  panel data,'' \emph{The Review of Economics and Statistics}, vol.~89, no.~2,
  pp. 221--233, 2007.

\bibitem[Conley and Taber(2011)]{conley2011inference}
T.~G. Conley and C.~R. Taber, ``Inference with ``difference in differences''
  with a small number of policy changes,'' \emph{The Review of Economics and
  Statistics}, vol.~93, no.~1, pp. 113--125, 2011.

\bibitem[Abadie et~al.(2010)Abadie, Diamond, and
  Hainmueller]{abadie2010synthetic}
A.~Abadie, A.~Diamond, and J.~Hainmueller, ``Synthetic control methods for
  comparative case studies: Estimating the effect of {C}alifornia's tobacco
  control program,'' \emph{Journal of the American Statistical Association},
  vol. 105, no. 490, pp. 493--505, 2010.

\bibitem[Crifasi et~al.(2018)Crifasi, Merrill-Francis, McCourt, Vernick,
  Wintemute, and Webster]{crifasi2018association}
C.~K. Crifasi, M.~Merrill-Francis, A.~McCourt, J.~S. Vernick, G.~J. Wintemute,
  and D.~W. Webster, ``Association between firearm laws and homicide in urban
  counties,'' \emph{Journal of urban health}, vol.~95, no.~3, pp. 383--390,
  2018.

\bibitem[Campbell and Stanley(1963)]{campbell1963experimental}
D.~T. Campbell and J.~C. Stanley, ``Experimental and quasi-experimental designs
  for research,'' in \emph{Handbook of research on teaching}, N.~Gage,
  Ed.\hskip 1em plus 0.5em minus 0.4em\relax Chicago, IL: Rand McNally, 1963,
  pp. 171--246.

\bibitem[Hasegawa et~al.(2017)Hasegawa, Deshpande, Small, and
  Rosenbaum]{hasegawa2017}
R.~Hasegawa, S.~Deshpande, D.~Small, and P.~Rosenbaum, ``Causal inference with
  two versions of treatment,'' 2017, posted on arXiv,
  https://arxiv.org/pdf/1705.03918.pdf.

\bibitem[Rudolph et~al.(2015)Rudolph, Stuart, Vernick, and
  Webster]{rudolph2015association}
K.~E. Rudolph, E.~A. Stuart, J.~S. Vernick, and D.~W. Webster, ``Association
  between connecticut's permit-to-purchase handgun law and homicides,''
  \emph{American journal of public health}, vol. 105, no.~8, pp. e49--e54,
  2015.

\bibitem[Imbens and Wooldridge(2009)]{imbens2009recent}
G.~W. Imbens and J.~M. Wooldridge, ``Recent developments in the econometrics of
  program evaluation,'' \emph{Journal of Economic Literature}, vol.~47, no.~1,
  pp. 5--86, 2009.

\bibitem[Hansen(2007)]{hansen2007generalized}
C.~B. Hansen, ``Generalized least squares inference in panel and multilevel
  models with serial correlation and fixed effects,'' \emph{Journal of
  Econometrics}, vol. 140, no.~2, pp. 670--694, 2007.

\bibitem[Blitstein et~al.(2005)Blitstein, Hannan, Murray, and
  Shadish]{blitstein2005increasing}
J.~L. Blitstein, P.~J. Hannan, D.~M. Murray, and W.~R. Shadish, ``Increasing
  the degrees of freedom in existing group randomized trials: The df*
  approach,'' \emph{Evaluation Review}, vol.~29, no.~3, pp. 241--267, 2005.

\bibitem[Lehmann(1952)]{lehmann1952testing}
E.~Lehmann, ``Testing multiparameter hypotheses,'' \emph{The Annals of
  Mathematical Statistics}, pp. 541--552, 1952.

\bibitem[Berger(1982)]{berger1982multiparameter}
R.~L. Berger, ``Multiparameter hypothesis testing and acceptance sampling,''
  \emph{Technometrics}, vol.~24, no.~4, pp. 295--300, 1982.

\end{thebibliography}

\newpage

\end{document}